\documentclass[lettersize,journal]{IEEEtran}
\usepackage{graphicx} 
\usepackage{amsmath,amsfonts}
\usepackage{algorithm}
\usepackage{algpseudocode}
\usepackage{array}
\usepackage[caption=false,font=normalsize,labelfont=sf,textfont=sf]{subfig}
\usepackage{textcomp}
\usepackage{stfloats}
\usepackage{url}
\usepackage{verbatim}
\usepackage{graphicx}
\usepackage{cite}
\usepackage{dsfont}
\usepackage{xcolor}
\usepackage{subfig}
\title{A Practical Validation of RIS Detection and Identification}

\author{Recep Vural, Aymen Khaleel,~\IEEEmembership{Member,~IEEE}, and Ertugrul Basar,~\IEEEmembership{Fellow,~IEEE}
\vspace{-1 cm}

\thanks{Copyright \copyright~2025 IEEE. Personal use of this material is permitted. However, permission to use this material for any other purposes must be obtained from the IEEE by sending a request to pubs-permissions@ieee.org.

Recep Vural is with the Communications Research and Innovation Laboratory
(CoreLab), Department of Electrical and Electronics Engineering, Ko\c{c}
University, 34450 Sariyer, Istanbul, Turkey. (e-mail: rvural22@ku.edu.tr).

Aymen Khaleel was with the Department of Electrical Engineering, Koc University during this work and is currently with the Faculty of Electrical Engineering and Information Technology, Ruhr-University Bochum, Bochum 44801, Germany. (e-mail: aymen.khaleel@ruhr-uni-bochum.de).
 
Ertugrul Basar was with the Department of Electrical Engineering, Koc University during this work and is currently with the Department of Electrical Engineering, Tampere University, 33720 Tampere, Finland (email: ertugrul.basar@tuni.fi).

{

}
}

}

\begin{document}
\maketitle

\begin{abstract} 
Reconfigurable intelligent surface (RIS)-assisted communication is a key enabling technology for next-generation wireless communication networks, allowing for the reshaping of wireless channels without requiring traditional radio frequency (RF) active components. While their passive nature makes RISs highly attractive, it also presents a challenge: RISs cannot actively identify themselves to user equipments (UEs). Recently, a new method has been proposed to detect and identify RISs by letting them modulate their identities in the signals reflected from their surfaces. In this letter, we first propose a new and simpler modulation method for RISs and then validate the concept of RIS detection and identification (RIS-ID) using a real-world experimental setup. The obtained results validate the RIS-ID concept and show the effectiveness of our proposed modulation method over different operating scenarios and systems settings. 

\end{abstract}


\vspace{-0.1 cm}
\section{Introduction}

\IEEEPARstart{R}{econfigurable} intelligent surfaces (RISs), composed of large arrays of controllable elements, are rapidly gaining prominence as an enabling technology for next-generation wireless communication systems, particularly in the context of sixth-generation (6G) networks \cite{RISfor6G}. By controlling the amplitude, phase, and polarization of the signals hitting their surfaces, RISs reshape the propagation environment and bring several advantages, such as extended coverage, enhanced data rates, and increased energy efficiency \cite{RIS_Advantages}. Moreover, RISs can facilitate better interference management \cite{RIS_Interference}, provide robust security through signal manipulation \cite{Security_RIS}, and enable precise localization for user equipment (UE) by altering the signal propagation paths\cite{RIS_Localization}. In parallel with theoretical developments, practical implementations of RISs have also progressed significantly, demonstrating their feasibility in real-world wireless systems \cite{2BitRISPrototyping,SpaceTimeDigitalCoding}.

While RISs are most commonly used as passive beamforming arrays in wireless systems to control the propagation environment, they have also been utilized to transmit additional information bits through index modulation (IM) schemes\cite{Chaotic_IM,KMeans_IM}.
Taking a different approach, several works have explored their potential to be used as an over-the-air modulator in wireless transmission systems\cite{LargeIntelligent}. In \cite{8PSKTransmitter}, an RIS-based 8-phase shift-keying (PSK) transmitter for single-input-single-output (SISO) transmission is proposed, which modulates the incident signal by controlling the phase response of the RIS elements. This approach achieves data transmission at a rate of $6.144$ Mb/s without the need for traditional radio frequency (RF) components. The authors of \cite{16AQMTransmitterMIMO} extended the idea to a multiple-input-multiple-output (MIMO) systems and developed an RIS-based RF chain-free 16-quadrature amplitude modulation (QAM) transmitter, achieving a data rate of $20$ Mb/s. 
Another notable application of RIS-based modulation is studied in \cite{LocalizationByModulatedRIS}, where RISs are utilized for UE localization. In this system, the base station (BS) transmits an unmodulated carrier signal, which is then modulated by multiple RISs at different time slots, with each RIS imparting a unique m-sequence. These uniquely assigned m-sequences enable the UE to differentiate received signals from different RISs. Finally, the location of the UE is estimated by applying time-difference-of-arrival (TDOA) method on the received signals. 

However, existing studies often presume that UEs are informed about the presence of deployed RISs, which is a reasonable assumption when the network provides this information to the UE. Nevertheless, the existence of RISs nearby, reported to the UE by the network, does not guarantee the establishment of a communication link due to potential obstacles between the RIS and the UE \cite{RIS_ID_Theoretical}. Therefore, to use an RIS effectively, it is essential first to determine whether the RIS is reachable by the UE. To the best of the authors' knowledge, this problem was first addressed in the literature in \cite{RIS_ID_Theoretical}, where the authors proposed a method called RIS-ID to enable UE to identify reachable RISs in its environment.
The RIS-ID method utilizes binary phase shift keying (BPSK) modulation by leveraging phase changes at the RIS. This method allowed the embedding of unique phase shift reflection patterns (PSRPs) into the carrier signal impinging on the RIS surfaces, enabling the UE to detect and uniquely identify reachable RISs. 

However, in \cite{RIS_ID_Theoretical}, since the RIS detection process is performed before synchronization, reflected BPSK symbols are expected to be distorted due to time, frequency, and phase offsets at the UE side, negatively affecting the correlation amplitude. In this letter, we address this limitation by proposing a new modulation method that is less sensitive to synchronization issues, particularly phase and frequency offsets. Our approach leverages the phase-dependent amplitude variations associated with RIS elements to mimic amplitude shift keying (ASK) modulation. Furthermore, unlike \cite{RIS_ID_Theoretical}, we validate the RIS-ID concept through a real-world experimental setup incorporating a variety of operating scenarios and system configurations.

The remainder of the letter is organized as follows. Section II presents the system model used for the proposed method and the detection algorithm that was developed to determine RIS reachability. Section III outlines the experimental setup and implementation. Section IV provides the experimental results and performance evaluation. Finally, Section V concludes the letter.

\section{System Model}
Here, we briefly introduce the RIS-ID concept and then provide the system model related to our proposed modulation method, as follows.
\vspace{-0.25 cm}
\subsection{ RIS Detection and Identification Method}

As proposed in \cite{RIS_ID_Theoretical}, the RIS-ID process involves the UE continuously transmitting an omnidirectional, unmodulated carrier signal while simultaneously receiving the echoes. Reachable RISs in the vicinity of the UE reflect this signal, modulating it with their unique PSRPs, which are vectors of $1$s and $-1$s. The modulation is achieved by adjusting the phase of the RIS elements to either $0$ or $\pi$, depending on whether the corresponding PSRP symbol is $1$ or $-1$. The UE then correlates the received signal with locally generated versions of the PSRPs, and if the correlation exceeds a predefined threshold, the corresponding RIS is declared reachable.

\vspace{-0.25 cm}
\subsection{ Proposed Modulation Method}

\begin{figure}[t!]
    \centering
    \includegraphics[width=80mm]{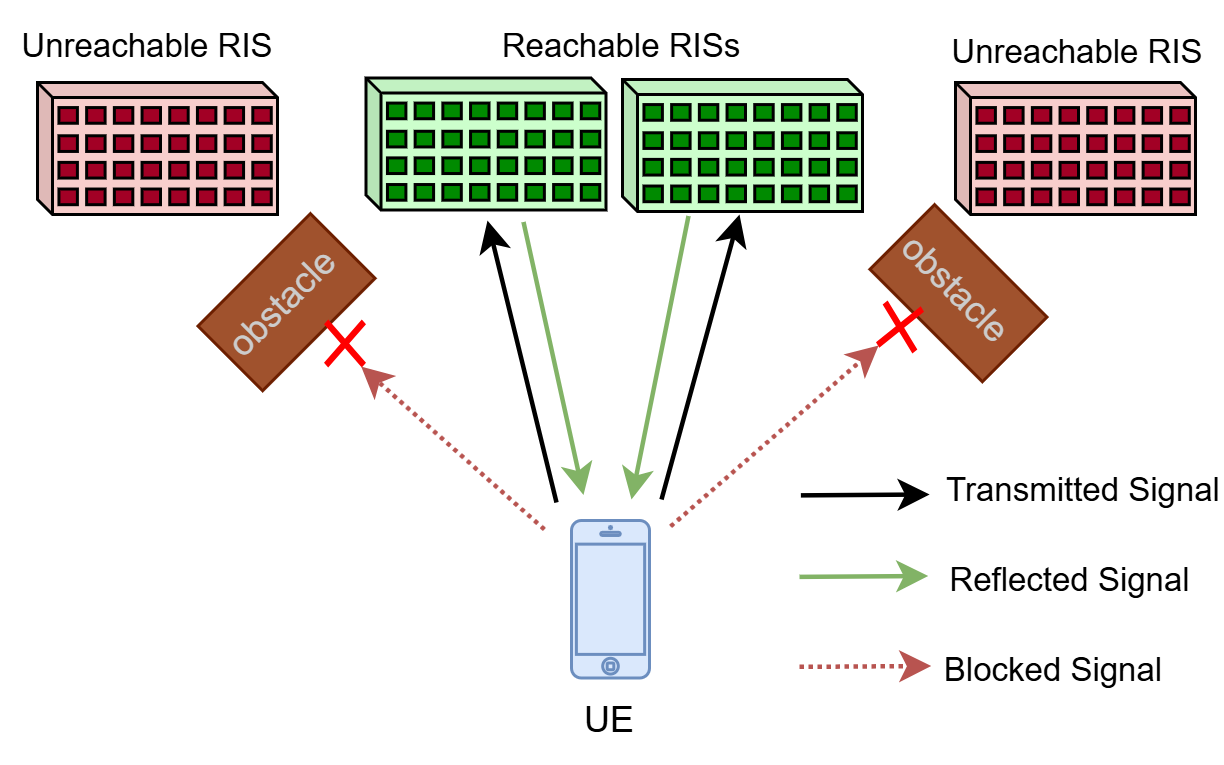}
    \caption{ A UE aims to identify nearby reachable RISs.}\label{fig:RISPracticalSystemModel}
\end{figure}

As illustrated in Fig. \ref{fig:RISPracticalSystemModel},  a UE seeks to identify reachable RISs in the environment using the signals reflected from their surfaces. Each RIS has an ID number uniquely associated with an amplitude reflection pattern (ARP), which is a vector of $1$'s and $-1$'s. In the identification process, UE continuously transmits an unmodulated sinusoidal carrier signal (corresponding complex baseband sample $x=1$) and simultaneously collects $M$ symbols, where $M$ denotes the length of the ARP vector. In contrast to \cite{RIS_ID_Theoretical}, where a unity amplitude reflection coefficient was assumed, and BPSK modulation was used, in this system, reachable RISs modulate the signal hitting their surface according to their ARPs by changing the amplitude of the reflected signals, exploiting the amplitude-phase correlation in RIS elements \cite{abeywickrama2020intelligent}.

In the proposed system, the UE and RISs are unsynchronized, and the detection process involves searching for the maximum correlation between the ARPs of RISs and the amplitude of the received symbols across all possible circular shifts. To ensure robust detection, the ARPs of the RISs must exhibit orthogonality to their circularly shifted versions. While Walsh-Hadamard (WH) codes are known for their optimal orthogonality in synchronous systems, their cross-correlation properties deteriorate in asynchronous settings due to timing misalignments. To address this, we select ARPs from a specific subset of WH codes whose circularly shifted versions maintain zero cross-correlation for all possible shifts, as detailed in Section \ref{Sec:ExpSetup}.

Consider $L$ possible reachable RISs in a wireless environment. Then, the $m$-th collected symbol at the UE can be given as\cite{RIS_ID_Theoretical}
\vspace{-0.2 cm}
\begin{align}
\label{eq:y_m}
y_m &=x\sum_{l=1}^{L}\left[(\mathbf{h}_{\text{UR},m}^{(l)})^T \mathbf{\Phi}_m^{(l)} \mathbf{h}_{\text{RU},m}^{(l)}\right]\mu^{(l)}+n_m \nonumber \\ \nonumber 
&\overset{\text{I}}=x\sum_{l=1}^{L}a_m^{(l)}e^{j\phi_{m}^{(l)}}\left[(\mathbf{h}_{\text{UR},m}^{(l)})^T  \mathds{I}_{N^{(l)}} \mathbf{h}_{\text{RU},m}^{(l)}\right]\mu^{(l)} +n_m\nonumber \\
 &\overset{\text{II}}=\sum_{l=1}^{L}a_m^{(l)}e^{j\phi_{m}^{(l)}}\tilde{h}^{(l)}\mu^{(l)}+n_m \nonumber\\
\end{align}
where $n_m\sim\mathcal{N}_\mathbb{C}(0,\sigma_n^2)$ is the additive white Gaussian noise (AWGN) sample; $\mathbf{h}_{\text{UR},m}^{(l)}\in\mathbb{C}^{N^{(l)}}$ and  $\mathbf{h}_{\text{RU},m}^{(l)}\in\mathbb{C}^{N^{(l)}} $ are UE-RIS and RIS-UE channel vectors, respectively, with $l$ denoting the RIS index and $N^{(l)}$ representing the size of the $l$-th RIS; $\mathbf{\Phi}_m^{(l)}$ denotes the RIS phase shift matrix; $\mu^{(l)} \in \{0,1\}$  represents the reachability of the RIS, where $\mu^{(l)}=1$ indicates $l$-th RIS is reachable, and $\mu^{(l)}=0$ indicates $l$-th RIS is unreachable. In step I, taking into account the ARP, we set $\mathbf{\Phi}_m^{(l)}=a_m^{(l)}e^{j\phi_{m}^{(l)}}\mathds{I}_{N^{(l)}}$ where $\mathds{I}_{N^{(l)}}$ is the identity matrix of size $N^{(l)}$, $\phi_{m}^{(l)}$ is the phase shift applied by RIS elements and $a_m^{(l)}$ is its amplitude coefficient (which depends on $\phi_{m}^{(l)}$). The relationship between amplitude coefficient $a$ and phase $\phi$ can be given as \cite{abeywickrama2020intelligent} 
\begin{align}
a(\phi) = \left(1 - a_{\text{min}}\right) \left(\frac{\sin\left(\phi - \delta-\frac{\pi}{2} \right) + 1}{2}\right)^\gamma + a_{\text{min}},
\label{eq:a_theta}
\end{align}
where \(a_{\text{min}} \geq 0\), \(\delta \geq 0\), and \(\gamma \geq 0\) are the constants related to the specific circuit implementation.\footnote{ The model parameters \(a_{\text{min}} \), \(\delta \), and \(\gamma\) in \eqref{eq:a_theta} may vary from one RIS to another RIS due to the difference in the circuit implementation of the RIS elements. However, \eqref{eq:a_theta} serves as a general phase-dependent amplitude variations model.}   \(a_{\text{min}}\) is the minimum amplitude coefficient, $\delta$ is the phase where the amplitude coefficient reaches its minimum value such that $a(\delta)=a_{\text{min}}$ for $\phi=\delta $ , and $\gamma$ controls the steepness of the function curve \cite{abeywickrama2020intelligent}. The RIS elements can apply two distinct phase shifts in our experimental setup, which will be detailed in Section \ref{Sec:ExpSetup}. Using \eqref{eq:a_theta}, corresponding to these two phase shifts, $a_m^{(l)}$ can take on two values: $\hat{a}_1$ and $\hat{a}_2$.

Furthermore, in step II of \eqref{eq:y_m}, assuming that the channel coherence time exceeds the duration required to receive all reflected symbols at the UE, we eliminate the symbol index $m$ and define $\tilde{h}^{(l)}=(\mathbf{h}_{\text{UR}}^{(l)})^T\mathds{I}_{N^{(l)}} \mathbf{h}_{\text{RU}}^{(l)}$.

Considering \eqref{eq:y_m} and \eqref{eq:a_theta}, the amplitude of the received signal can be manipulated by changing $a_m^{(l)}$. Accordingly, the RIS induces its ARP by mapping ARP symbols to their corresponding amplitude coefficient values $\hat{a}_1$ or $\hat{a}_2$. Specifically, at the reflection of the $m$-th symbol, $a_m^{(l)}$ is defined as 
\begin{align}
a_m^{(l)}=
\begin{cases} 
      \hat{a}_1, & q_{\tilde{m}}^{(l)}=1 \\
      \hat{a}_2, & q_{\tilde{m}}^{(l)}=-1
   \end{cases}
   ,\;m,\tilde{m}\in\{1,2,\dots,M\},
\end{align}
where $q_{\tilde{m}}^{(l)} \in \{-1, 1\}$ is the $\tilde{m}$-th element of the ARP vector $\mathbf{q}^{(l)}=[q_1^{(l)},\cdots,q_M^{(l)}]$ with $\tilde{m}=\left((c+m-1)\mod M\right) +1$ denoting the index of the $m$-th ARP symbol modulating $x$. Here, $c$ is the index of the first ARP symbol modulating $x$, modeled as a uniformly distributed random integer $c\in\mathcal{U}\{1,2,\dots,M\}$. 

\begin{algorithm}[t!]
    \caption{ RIS-ID Algorithm }
    \textbf{Input:} $ l, \mathbf{q}^{(l)}, M, thr $ \\
    \textbf{Output:} $\widehat{\mu}^{(l)}$ 
    \begin{algorithmic}[1]
        \State Obtain the amplitude of the received symbols to form $\tilde{\mathbf{y}}=[\tilde{y}_{1}, \tilde{y}_{2},\dots, \tilde{y}_{M}].$ 
        
        \State $D^{(l)}_{\text{max}} \leftarrow 0$
        \For{$i = 1$ to $M$}
            \State $\mathbf{q}^{(l)}_i \leftarrow \text{circshift}(\mathbf{q}^{(l)}, i)$   : circular shift 
            \State $d_i^{(l)} \leftarrow \frac{\mathbf{q}^{(l)}_i \cdot \tilde{\mathbf{y}}}{\sqrt{M}}$ 
            \State $D_i^{(l)} \leftarrow \left|d_i^{(l)}\right|^2$
            \If{$D_i^{(l)} > D^{(l)}_{\text{max}}$} 
                \State $D^{(l)}_{\text{max}} \leftarrow D_i^{(l)}$ 
            \EndIf
        \EndFor
        \If{$D^{(l)}_{\text{max}} > thr$}
            \State $\widehat{\mu}^{(l)} \leftarrow 1$
        \Else
            \State $\widehat{\mu}^{(l)} \leftarrow 0$
        \EndIf
        \State \textbf{return} $\widehat{\mu}^{(l)}$
    \end{algorithmic}
\end{algorithm}

Consequently, the amplitude of the received vector \( \mathbf{y}=[y_{1}, y_{2},\dots, y_{M}] \) relates to the cyclically shifted version of \( \mathbf{q}^{(l)} \) from each reachable RIS. By quantifying these relationships, UE aims to detect reachable RISs in the environment.

The designed detection process is outlined in Algorithm 1, which takes as input $l$, $\mathbf{q}^{(l)}$, $M$, and a predefined detection threshold $thr$. The output is the reachability estimate of $l$-th RIS $\hat{\mu}^{(l)}$. The algorithm begins by obtaining the vector $\tilde{\mathbf{y}}=[\tilde{y}_{1}, \tilde{y}_{2},\dots, \tilde{y}_{M}]$, which contains the amplitude of the received symbols. Next, the algorithm iteratively circularly shifts the ARP vector $\mathbf{q}^{(l)}$ by $i$ positions, where $i$ ranges from 1 to $M$. This shifting is necessary due to the unsynchronization between the RIS and UE, which causes the received signal to be modulated according to a cyclically shifted version of the ARP \cite{RIS_ID_Theoretical}. For each shift, the algorithm calculates $d_i^{(l)}$ by correlating the shifted ARP vector $\mathbf{q}^{(l)}_i$ with $\tilde{\mathbf{y}}$ and normalizes it by $\sqrt{M}$. The detection value $D_i^{(l)}$ is then obtained as the squared magnitude of $d_i^{(l)}$. The algorithm tracks the maximum detection value, $D_{\text{max}}^{(l)}$, across all possible shifts and compares it with $thr$. If \( D_{\text{max}}^{(l)} \) exceeds $thr$, the RIS is considered reachable, and \( \hat{\mu}^{(l)} \) is set to 1; otherwise, it is considered unreachable, and \( \hat{\mu}^{(l)} \) is set to 0. By repeating Algorithm 1 for $ l = 1, 2, \dots, L$, the reachability estimate of each RIS is obtained.

As in \cite{RIS_ID_Theoretical}, the performance of the proposed method can be assessed by false detection and miss detection probabilities. The false detection probability $P_F$ indicates the probability of declaring an RIS is reachable while it is unreachable by UE. Conversely, the miss detection probability $P_\text{miss}$ indicates the probability of declaration of an RIS as unreachable while it is reachable. These probabilities can be mathematically expressed as  
$P_F= P(\widehat{\mu}^{(l)}=1|\mu^{(l)}=0)$ and $P_\text{miss}=P(\widehat{\mu}^{(l)}=0|\mu^{(l)}=1)$. 

\label{Sec:ExpSetup}
\begin{figure}[t!]
    \centering
    \includegraphics[width=87mm]{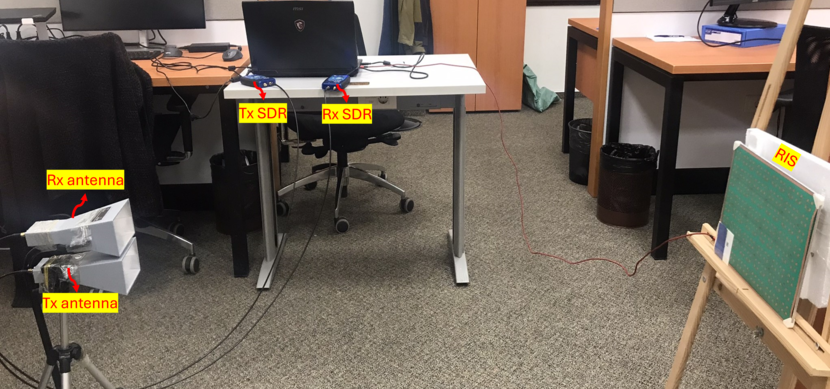}
    \caption{ The experiment setup to validate the RIS-ID scheme. }\label{fig:RIS ID PracticalSetup}
    \vspace{-0.2 cm}
\end{figure}

\vspace{-0.2 cm}

\section{Experimental Setup and Implementation}
The measurement setup used in the experiment is shown in Fig. \ref{fig:RIS ID PracticalSetup}. The setup consists of an RIS hardware designed by Greenerwave and composed of 76 reflecting elements. Each element is equipped with two PIN diodes to control the phase shift applied to the incident signal in both vertical and horizontal polarization\cite{RISE6G2022}. For the transmission and reception, ADALM-PLUTO software-defined radio (SDR) modules as  well as horn antennas are used. An unmodulated signal at a frequency of $5.27$ GHz is transmitted from transmitter (Tx) SDR using a horn antenna directed at the RIS, which is located in the far field of the antennas. The RIS is controlled by a computer to continuously change its phase shift matrix, creating unique ARPs. The signal reflected back from the RIS surface is received by the horn antenna of the receiver (Rx) SDR to be used in the detection process described in Algorithm 1.

In this experiment, we used two possibly reachable RISs ($ L = 2 $) in the environment: RIS 1 and RIS 2. Specifically, the experiment is carried out for the following scenarios: 'RIS 1 Reachable' (${\mu}^{(1)}=1$, ${\mu}^{(2)}=0$), 'Both RISs Reachable' (${\mu}^{(1)}=1$, ${\mu}^{(2)}=1$) and 'No RIS Reachable' (${\mu}^{(1)}=0$, ${\mu}^{(2)}=0$). Due to the limitation of having only one RIS hardware available in our experimental setup, the 'Both RISs Reachable' scenario was realized by virtually partitioning the RIS into two separate segments, representing RIS 1 and RIS 2. For each scenario, the experiment was performed under various settings of \( M \) and the Tx gain,\footnote{Note that the ADALM-Pluto SDR does not provide a direct parameter for setting the Tx power. Instead, it offers control over the Tx gain, which serves as an indirect way for adjusting the output power.} where 300 measurements are taken for each setting. 
The performance of the proposed method was then evaluated based on $ P_{\text{miss}} $, $ P_F $, and the average maximum detection value, $ \overline{D^{(l)}_{\text{max}}} = \frac{1}{300} \sum_{k=1}^{300} D^{(l)}_{\text{max},k} $, where $ k $ denotes the measurement index. 

In the experiment, the ARP vectors of the RISs for $M = 8$ are selected from WH codes as: $\mathbf{q}^{(1)}=[1,-1,1,-1,1,-1,1,-1]$ and  $\mathbf{q}^{(2)}=[1,1,-1,-1,-1,-1,1,1]$. To extend the ARPs to longer lengths, we leverage the property that concatenating a WH code with itself produces another valid WH code. For instance, the ARP vector for \( M = 16 \) is obtained by concatenating the \( M = 8 \) vector with itself. Similarly, the ARP vector for \( M = 32 \) is obtained by concatenating two \( M = 16 \) vectors.

\begin{figure}[t!]
    \centering
    \includegraphics[width=70mm]{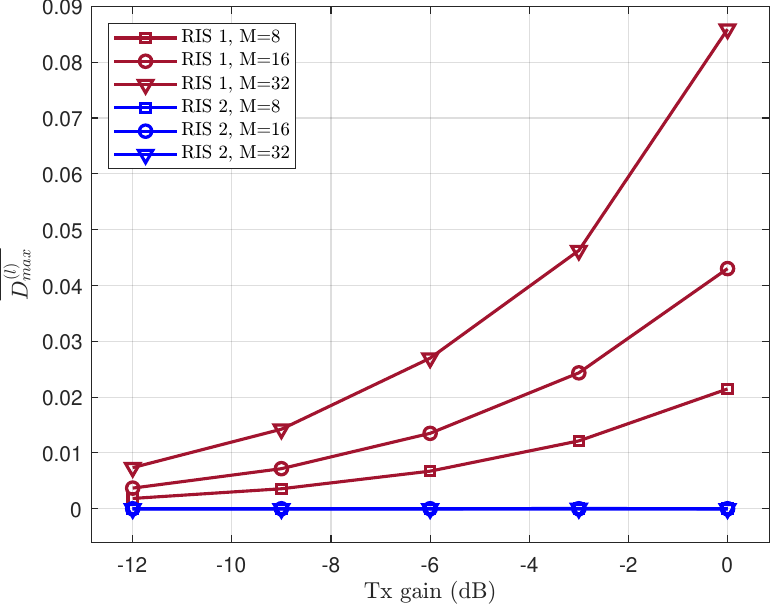}
    \caption{ The effect of Tx gain and ARP length $M$ on $\overline{D^{(l)}_{\text{max}}}$ for RIS 1 and RIS 2 in "RIS 1 Reachable" scenario.}\label{fig:Reachable_RIS1_Correlation_ForM}

\end{figure}

\vspace{-0.2 cm}
\section{Experimental Results}

In this section, we provide experimental results to assess the performance of the practical RIS-ID method. In the experiments, unless otherwise specified, the code length is set as $M=16$.  

\vspace{-0.3 cm}
\subsection{RIS 1 Reachable Scenario}

In this scenario, all reflecting elements are assigned to RIS 1. Accordingly, the states of elements are adjusted according to the ARP of RIS 1. As seen in Fig. \ref{fig:Reachable_RIS1_Correlation_ForM}, when only RIS 1 is reachable, $\overline{D^{(2)}_{\text{max}}}$ has low value and does not increase with the increase in Tx gain or ARP length $M$. On the other hand, $\overline{D^{(1)}_{\text{max}}}$ is consistently higher than $\overline{D^{(2)}_{\text{max}}}$, and increases with both the Tx gain and $M$. For example, increasing the Tx gain by $3$ dB approximately doubles $\overline{D^{(1)}_{\text{max}}}$ and, doubling $M$ increases the $\overline{D^{(1)}_{\text{max}}}$ by a factor of two, due to the increase in the correlation amplitude.

\begin{figure}[t!]
    \centering
    \includegraphics[width=70mm]{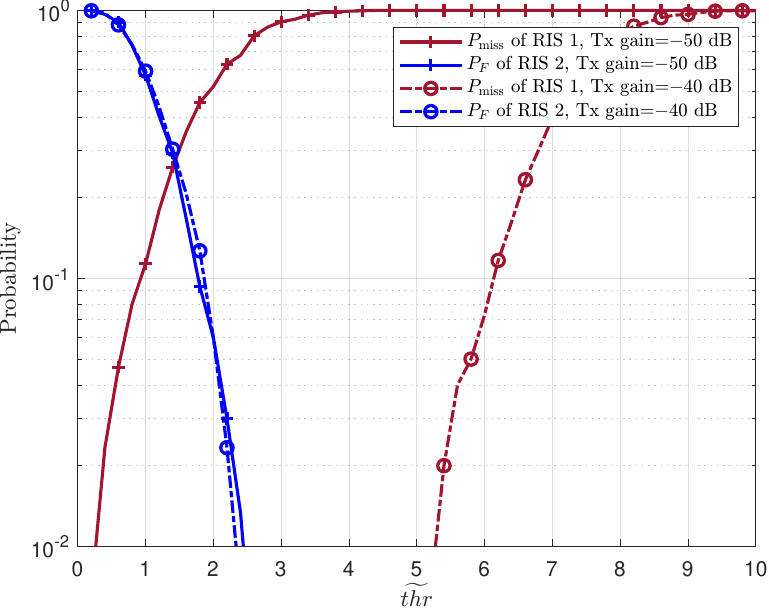}
    \caption{The effect of the Tx gain on $P_{miss}$ of RIS 1 and $P_{F}$ of RIS 2 in "RIS 1 Reachable" scenario.}\label{fig:MissDetection_RIS1_FalseDetection_RIS2}
    \vspace{-0.1 cm}
\end{figure}

In Fig. \ref{fig:MissDetection_RIS1_FalseDetection_RIS2}, the effect of the Tx gain on $P_{miss}$ of RIS 1 and $P_{F}$ of RIS 2 is examined, with the \textit{x}-axis of the figure representing the normalized detection threshold $\widetilde{thr}$. Here, $\widetilde{thr}=thr/P_{noise}$, where $P_{noise}$ is the noise power measured at the Rx SDR. From Fig. \ref{fig:MissDetection_RIS1_FalseDetection_RIS2}, it is observed that while increasing the Tx gain decreases $P_{miss}$ of RIS 1, it does not affect $P_{F}$ of RIS 2 considerably. Additionally, it is clearly seen that increasing $\widetilde{thr}$ decreases $P_{F}$ of RIS 2 while increases $P_{miss}$ of RIS 1 as expected \cite{RIS_ID_Theoretical}.

\begin{figure}[!t]
    \centering
    \includegraphics[width=70mm]{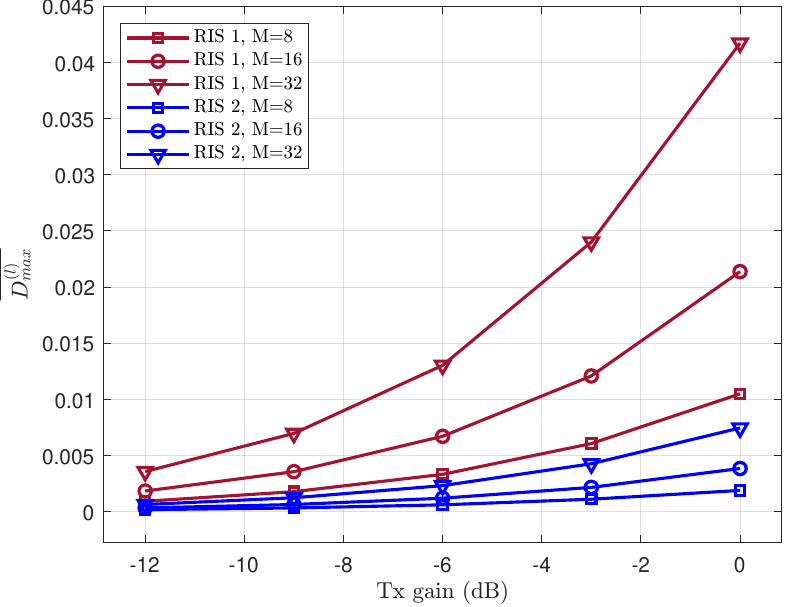}
    \caption{ The impact of $M$ on $\overline{D^{(l)}_{\text{max}}}$ for RIS 1 and RIS 2 in "Both RISs reachable" scenario.}\label{fig:Reachable_BothRISs_Correlation_ForM}
\end{figure}

\vspace{-0.45 cm }
\subsection{Both RISs Reachable Scenario}

In this scenario, while approximately half of reflecting elements are assigned to RIS 1, the other half are assigned to RIS 2. Accordingly, the states of elements are adjusted based on their assignments: those assigned to RIS 1 are configured according to the ARP of RIS 1, while those assigned to RIS 2 are configured according to the ARP of RIS 2. In Fig. \ref{fig:Reachable_BothRISs_Correlation_ForM}, we show the effect of $M$ and Tx gain on $\overline{D^{(l)}_{\text{max}}}$ for RIS 1 and RIS 2 when both RISs are reachable. It is observed that $\overline{D^{(1)}_{\text{max}}}$ is higher than $\overline{D^{(2)}_{\text{max}}}$. This behavior is resulted from that the transmitted signal can not be focused on both RISs fairly due to the miss alignment between the different components of the experiment. Additionally, it is seen that increasing the Tx gain increases $\overline{D^{(l)}_{\text{max}}}$ for both RISs where increasing the Tx gain $3$ dB leads to approximately two-fold increase in $\overline{D^{(l)}_{\text{max}}}$ values.
Similarly, increasing $M$ also increases $\overline{D^{(l)}_{\text{max}}}$ for both RISs, where doubling $M$ doubles $\overline{D^{(l)}_{\text{max}}}$.

\begin{figure}[!t]
    \centering
    \includegraphics[width=70mm]{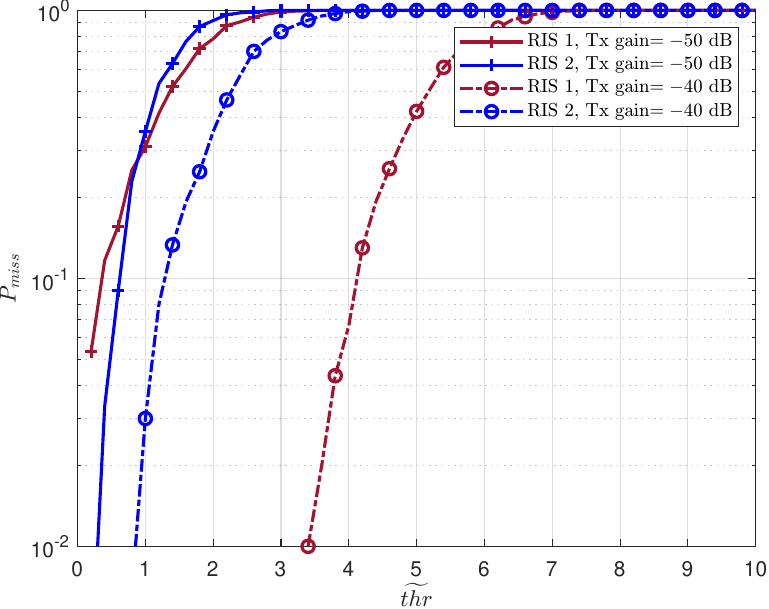}
    \caption{ $P_{miss}$ vs $\widetilde{thr}$ with different Tx gain values in "Both RISs reachable" scenario.}\label{fig:MissDetection_RIS1RIS2}
\end{figure}

In Fig. \ref{fig:MissDetection_RIS1RIS2}, the effect of $\widetilde{thr}$ on $P_{miss}$ of RIS 1 and RIS 2 is investigated. The results show that increasing $\widetilde{thr}$ leads to a higher $P_{miss}$ for both RISs. Specifically, when the Tx gain is $-50$ dB, increasing $\widetilde{thr}$ from $0.4$ to $1.2$ increases $P_{miss}$ from $0.12$ to $0.41$ for RIS 1 and from $0.03$ to $0.53$ for RIS 2. Moreover, increasing the Tx gain decreases the $P_{miss}$ for both RISs. This behavior is evident from Fig. \ref{fig:MissDetection_RIS1RIS2} for RIS 2. For example, when $\widetilde{thr}$ is set at $1$, increasing the Tx gain by $10$ dB reduces the $P_{miss}$ for RIS 2 from $0.35$ to $0.03$.

\vspace{-0.25 cm}
\subsection{No RIS Reachable Scenario}

\begin{figure}[t!]
    \centering
    \includegraphics[width=70mm]{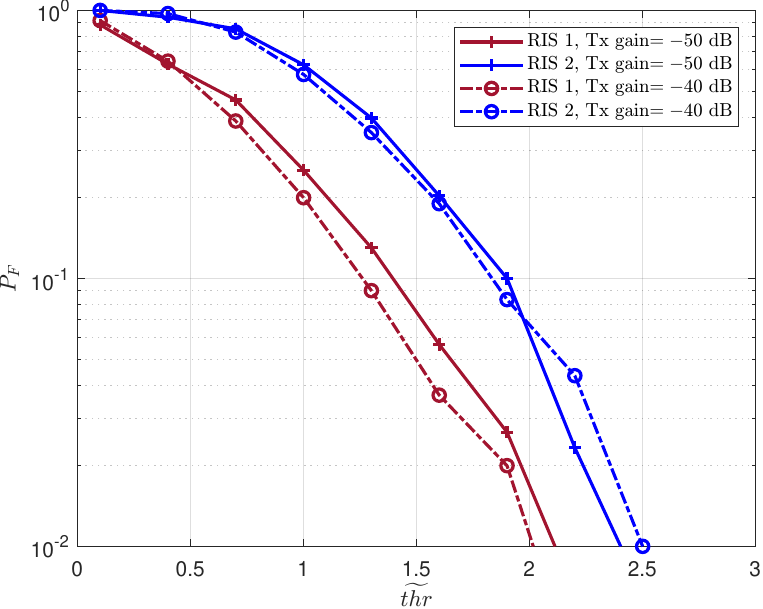}
    \caption{ $P_{F}$ vs $\widetilde{thr}$ with different Tx gain values in "No RIS reachable" scenario.}\label{fig:FalseDetection_RIS1RIS2}
\end{figure}

In this scenario, the RIS is removed from the environment, and the identification process is performed using Algorithm 1. Then, the effect of $\widetilde{thr}$ on $P_F$ of RIS 1 and RIS 2 is investigated. As seen in Fig. \ref{fig:FalseDetection_RIS1RIS2}, increasing $\widetilde{thr}$ decreased $P_F$ of RIS 1 and RIS 2. For example, when the Tx gain is $-50$ dB, increasing $\widetilde{thr}$ from $1$ to $1.6$, decreased $P_F$ of RIS 1 from $0.25$ to $0.06$ and decreased $P_F$ of RIS 2 from $0.63$ to $0.20$.  

Finally, we note that although the effect of the overall Tx–RIS–Rx link distance on the performance metrics was not explicitly investigated in any scenarios of the experiment, it can be inferred from the results in which Tx gain was varied. Increasing the Tx gain enhances the signal-to-noise ratio (SNR) at the receiver and thereby increases $\overline{D^{(l)}_{\text{max}}}$ for reachable RISs, as shown in Figs.~\ref{fig:Reachable_RIS1_Correlation_ForM} and \ref{fig:Reachable_BothRISs_Correlation_ForM}. This increase makes it more likely for $\overline{D^{(l)}_{\text{max}}}$ to exceed the detection threshold for reachable RISs. As a result, a higher Tx gain leads to a lower $P_{miss}$, while the $P_F$ remains mostly unchanged, as observed in Figs.~\ref{fig:MissDetection_RIS1_FalseDetection_RIS2}, \ref{fig:MissDetection_RIS1RIS2}, and \ref{fig:FalseDetection_RIS1RIS2}. On the other hand, increasing the Tx–RIS–Rx distance would introduce higher path loss, which in turn reduces the SNR at the receiver and consequently decreases $\overline{D^{(l)}_{\text{max}}}$ for reachable RISs. Therefore, increasing the Tx–RIS–Rx distance is analogous to decreasing the Tx gain and would be expected to increase $P_{miss}$ of reachable RISs while not changing $P_F$ of unreachable RISs considerably.

\vspace{-0.1cm}
\section{Conclusion}
This letter has proposed a novel modulation method for the RIS identification (RIS-ID) process to identify reachable RISs with less sensitivity to synchronization issues such as phase and frequency offsets. The proposed method has been tested through practical experiments using RIS hardware and software-defined radio (SDR) modules. Experimental results, including false and miss detection probability metrics, demonstrated the effectiveness of the proposed method in various scenarios with different system settings. Future work will explore optimizing the method for dynamic environments and integrating it with other network functions like beamforming and localization.
\vspace{-0.2 cm}

\bibliographystyle{IEEEtran}
\bibliography{refs}

\end{document}